\documentclass[a4paper,10pt,twoside]{cpc-hepnp}

\usepackage{multicol}
\usepackage{graphicx}
\usepackage{booktabs}
\usepackage{amssymb,bm,mathrsfs,bbm,amscd}
\usepackage[tbtags]{amsmath}
\usepackage{lastpage}

\begin{document}

\fancyhead[c]{\small Chinese Physics C~~~Vol. **, No. * (2014)
010201} \fancyfoot[C]{\small 010201-\thepage}

\footnotetext[0]{Received 24 April 2014}

\title{Spin-dependent $\gamma$ softness or triaxiality in even-even $^{132-138}$Nd nuclei
\thanks{Supported by National Natural Science Foundation of
China (10805040,11175217), Foundation and Advanced Technology
Research Program of Henan Province(132300410125), S \& T Research
Key Program of Henan Province Education Department (13A140667).}}

\author{%
      CHAI Qing-Zhen(²ñÇåìõ)$^{1}$%
\quad Wang Hua-Lei(Íõ»ªÀÚ)$^{1;1)}$\email{wanghualei@zzu.edu.cn}%
\quad Yang Qiong(ÑîÇí)$^{1}$ \quad LIU Min-Liang(ÁøÃôÁ¼)$^{2}$ }
\maketitle

\address{%
$^1$ School of Physics and Engineering, Zhengzhou University,
Zhengzhou 450001, China\\
$^2$ Institute of Modern Physics, Chinese Academy of Sciences,
Lanzhou 730000, China \\
}

\begin{abstract}
The properties of $\gamma$ instability in rapidly rotating even-even
$^{132-138}$Nd isotopes have been investigated using the
pairing-deformation self-consistent total-Routhian-surface
calculations in a deformation space of ($\beta_2, \gamma, \beta_4$).
It is found that even-even $^{134-138}$Nd nuclei exhibit the
triaxiality in both ground and excited states, even up to high-spin
ones. The lightest isotope possesses a well-deformed prolate shape
without $\gamma$ deformation component. The current numerical
results are compared with previous calculations and available
observables, showing basically a general agreement with the observed
trend of $\gamma$ correlations. The existing differences between
theory and experiment are analyzed and discussed briefly.

\end{abstract}



\footnotetext[0]{\hspace*{-3mm}\raisebox{0.3ex}{$\scriptstyle\copyright$}2013
Chinese Physical Society and the Institute of High Energy Physics of
the Chinese Academy of Sciences and the Institute
of Modern Physics of the Chinese Academy of Sciences and IOP Publishing Ltd}%

\begin{multicols}{2}

\section{Introduction}

Atomic nuclei exhibit a variety of shapes which are generally
sensitive to the single-particle structure, the collective behavior
and the total angular momentum. Study has revealed that  most of
deformed nuclei possess axially symmetric shapes (prolate or
oblate), which was confirmed by the observation of rotational band
structures and measurements of their properties~\cite{Tajima2001}.
Nevertheless, evidence for nonaxial $\gamma$ deformations (or
softness) has so far been widely found in nuclear spectroscopy. For
instance, some amazing characteristics caused possibly by the
$\gamma$ deformations, such as wobbling, signature inversion (or
splitting)and chiral doublets, were observed in many
nuclei~\cite{Bengtsson1984, Meng1997} . One expects that potential
energy surfaces that are $\gamma$ soft or which display deep minima
with nonzero $\gamma$ value would produce rather different nuclear
spectra, but it is not the case. Indeed, the question of whether
non-axially-symmetric nuclei are $\gamma$ soft or triaxial has been
an ongoing and active issue in nuclear structure physics for over
fifty years. It has been investigated extensively using theoretical
approaches that are essentially based on a rigid triaxial
potential~\cite{Davydov1958} and a completely $\gamma$-flat
($\gamma$-unstable) potential~\cite{Wilets1956}.  A further
discussion on signatures of $\gamma$ softness or triaxiality in low
energy nuclear spectra was performed about $20$ years ago by Zamfir
and Casten~\cite{Zamfir1991}. During the past several decades,
numerous studies have been carried out in terms of various
theoretical approaches including mean-field models and beyond
mean-field models. More recently, total-Routhian-surface (TRS)
calculations have been carried out for even-even germanium and
selenium isotopes to search for possible stable triaxial
deformations of nuclear states~\cite{Nomura2012}. The $\gamma$
softness in medium-heavy and heavy nuclei has been investigated in
the framework of energy density functionals~\cite{Casten1985}.

However, despite considerable effort, the precise description of
axially asymmetric shapes and the resulting triaxial quantum
many-body rotors still remain open problems. It is well known that
many of the nuclei in the $A = 130\thicksim140$ mass of transitional
region show the interesting characteristic feature known as the
triaxiality or a high degree of $\gamma$-softness, which arises from
the interplay of the valence protons and neutrons occupying
respectively low-lying and high-lying Nilsson orbitals within the
$h_{11/2}$ j-shell~\cite{Chen1983}. Their nuclear spectra usually
can be satisfactorily described using the corresponding $O(6)$
dynamical symmetry of the interacting boson model~\cite{Saito2008}.
In the present work, we perform TRS calculations with the inclusion
of the $\gamma$ deformation for several selected even-even
$^{132-138}$Nd isotopes in this mass region, focusing on their
evolutions of $\gamma$-softness or triaxiality with rotation and
providing a test for present model. We have investigated the
evolutions of octupole-softness in rotating $^{106,108}$Te and
neutron-deficient U isotopes using the similar TRS
calculations~\cite{Wang20121,Wang20122}. Experimentally, the
high-spin behaviors in even-even have been studied and interpreted
on the base of triaxial
shapes~\cite{Wadsworth1988,Angelis1994,Petrache1996,Petrache1999,Mukhopadhyay2008}.
Moreover, the quasi-$\gamma$ bands have been identified in these
nuclei~\cite{Saito2008,Paul1987,Kortelahti1990,Angelis1994,Petrache2012},
even the multiphonon $\gamma$-vibrational bands in
$^{138}$Nd~\cite{Li2013}.

\section{The theoretical framework}
The TRS calculation applied here is based on the
pairing-deformation-frequency self-consistent cranked shell model
(CSM)~\cite{Satula1994,Xu2000}. Such approach usually accounts well
for the overall systematics of high-spin phenomena in rapidly
rotating medium and heavy mass nuclei. The total Routhian, which is
called "Routhian" rather than "energy" in a rotating frame of
reference, is the sum of the energy of the non-rotating state and
the contribution due to cranking,
\begin{eqnarray}
\label{one} E^{\omega}(Z,N,\hat{\beta})& = &
 E^{\omega=0}(Z,N,\hat{\beta}) \nonumber\\[1mm]
 &&+[\langle\Psi{^\omega}\mid\hat{H}{^\omega}(Z,N,\hat{\beta})\mid\Psi{^\omega}\rangle \nonumber\\[1mm]
 &&-\langle\Psi{^\omega}\mid\hat{H}{^\omega}(Z,N,\hat{\beta})\mid\Psi{^\omega}\rangle^{\omega=0}].
\end{eqnarray}
The energy $E^{\omega=0}(Z,N,\hat{\beta})$ of the non-rotating state
consists of a macroscopic part, being a smooth function of Z, N and
deformation, and the fluctuating microscopic one, which is based on
some phenomenological single-particle potential, that is,
\begin{eqnarray}
E^{\omega=0}(Z,N,\hat{\beta}) &=& E_{macr}(Z,N,\hat{\beta}) +
E_{micr}(Z,N,\hat{\beta}).
\end{eqnarray}
where the macroscopic term is obtained from the sharp-surface
standard liquid-drop formula with the parameters of Myers and
Swiatecki~\cite{Myers1966}. The microscopic correction part, which
arises because of the non-uniform distribution of single-particle
levels in the nucleus, mainly contains a shell correction and a
pairing correction:
\begin{eqnarray}
E_{micr}(Z,N,\hat{\beta})  &=& \delta E_{shell}(Z,N,\hat{\beta})+
\delta E_{pair}(Z,N,\hat{\beta}).\nonumber\\[1mm]
\end{eqnarray}
These two contributions both can be evaluated from a set of
single-particle levels. Cranking indicates that the nuclear system
is constrained to rotate around a fixed axis (e.g., the $x-$axis) at
a given frequency $\omega$. This is equivalent to minimizing the
rotation hamiltonian $H^{\omega}=H-\omega I_x$ instead of the
hamiltonian $H$ with respect to variations of the mean field. For a
given rotational frequency and point of deformation lattice, this
can be achieved by solving the well known
Hartree-Fock-Bogolyubov-Cranking (HFBC) equations using a
sufficiently large space of single-particle states. Then one can
obtain the energy relative to the non-rotating state at $\omega =
0$, as mentioned in Eq.~(\ref{one}). After the numerical calculated
Routhians at fixed $\omega$ are interpolated using cubic spline
function between the lattice points, the equilibrium deformation can
be determined by minimizing the calculated TRS.

Note that nuclear shape is defined by the standard parametrization
in which it is expanded in spherical harmonics
$Y_{\lambda\mu}(\theta,\phi)$ ~\cite{Cwiok1987}. There is a
fundamental limitation in $\lambda$, because the range of the
individual "bumps" on the nuclear surface decreases with increasing
$\lambda$ and should not be smaller than a nucleon diameter
obviously~\cite{Greiner1996}. A limiting value of $\lambda<A^{1/3}$
can be obtained by a crude estimate~\cite{Greiner1996}. Therefore,
the deformation parameter $\hat{\beta}$ includes $\beta_2, \gamma$,
and $\beta_4$ where $\gamma$ describes triaxial shapes.
Single-particle energies needed above are obtained from a
phenomenological Woods-Saxon (WS) potential~\cite{Naza1989,
Cwiok1987} with the parameter set widely used for cranking
calculations. During the diagonalization process of the WS
hamiltonian, the deformed harmonic oscillator states with the
principal quantum number $N \leqslant 12$ and 14 have been used as a
basis for protons and neutrons, respectively. The shell and pairing
corrections at each deformation point are calculated by use of
Strutinsky method~\cite{Strutinsky1967} and Lipkin-Nogami (LN)
method~\cite{Pradhan1973}, respectively. The Strutinsky smoothing is
performed with a sixth-order Laguerre polynomial and a smoothing
range $\gamma =1.20\hbar \omega _0$, where $\hbar \omega
_0=41/A^{1/3}$ MeV. The LN method avoids the spurious pairing phase
transition encountered in the simpler BCS calculation. In the
pairing windows, empirically dozens of single-particle levels, the
respective some states (e.g.,half of the particle number $Z$ or $N$)
just below and above the Fermi energy, are included for both protons
and neutrons. Moreover, not only monopole but also doubly stretched
quadrupole pairings are considered. The monopole pairing strength,
$G$, is determined by the average gap method \cite{Moller1992} and the
quadrupole pairing ones are obtained by restoring the Galilean
invariance broken by the seniority pairing
force~\cite{Sakamoto1990}. Certainly, pairing correlations are
dependent on rotational frequency as well as deformation. During
solving the HFBC equations, pairing is treated self-consistently and
symmetries of the rotating potential are used to simplify the
cranking equations. In the reflection-symmetric case, both
signature, $r$, and intrinsic parity, $\pi$ are good quantum
numbers.

\section{Results and discussions}

The present TRS method, similar to most of the existing cranking
calculations, assumes that the rotational axis coincides with one of
the principal axis (the $x$ axis is generally chosen) of the
triaxial potential including $\beta_2, \gamma$, and $\beta_4$
deformations. In the actual calculations the Cartesian quadrupole
coordinates $X=\beta_2cos(\gamma+30^\circ)$ and
$Y=\beta_2sin(\gamma+30^\circ)$ were used, where the parameter
$\beta_2$ specifies the magnitude of the quadrupole deformation,
while $\gamma$ specifies the asymmetry of the shape. In the Lund
convention adopted here, the triaxiality parameter cover the range
$-120^\circ \leqslant \gamma \leqslant 60^\circ$ and the three
sectors $[-120^\circ, -60^\circ]$, $[-60^\circ, 0^\circ]$ and
$[0^\circ, 60^\circ]$ represent the same triaxial shapes but
represent rotation about the long, medium and short axes,
respectively. Certainly, for $\gamma=-120^\circ$ (prolate shape) the
nucleus rotates around the prolate symmetry axis and for
$\gamma=60^\circ$ (oblate shape) around the oblate symmetry axis.
For $\gamma=0^\circ$ (prolate shape) and for $\gamma=-60^\circ$
(oblate shape) the nucleus has a collective rotation around an axis
perpendicular to the symmetry axis.

\begin{center}
\includegraphics[width=8.5cm]{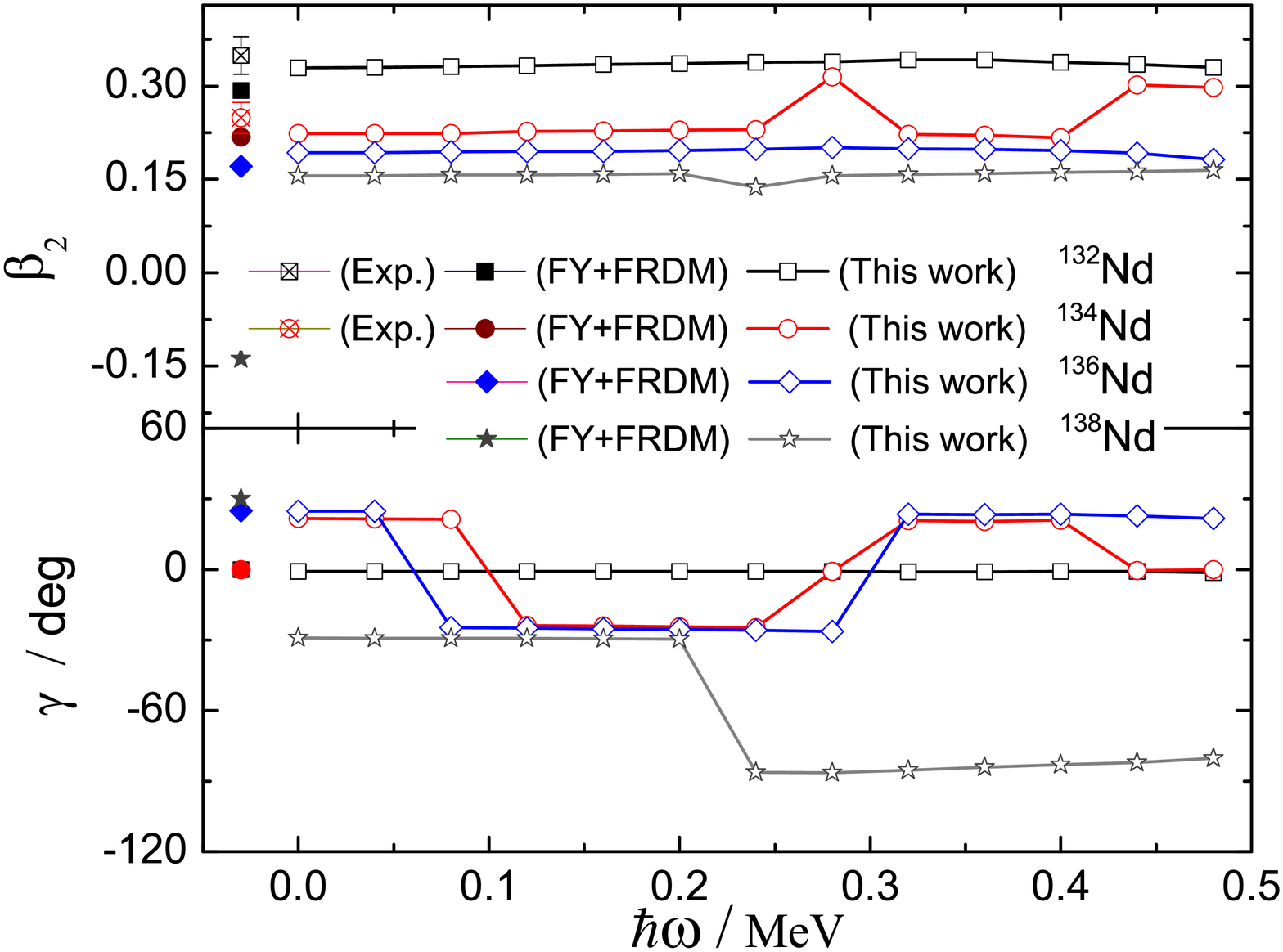}
\figcaption{\label{deformation} Calculated deformation parameters
$\beta_2$ (top) and $\gamma$ (bottom) of yrast states for even-even
nuclei $^{132-138}$Nd as a function of the rotational frequency
$\hbar \omega$, compared with the FY +FRDM
calculations~\cite{muller1995,muller2008} and partial experimental
values obtained from reduced transition probabilities B(E2) for the
ground states~\cite{Raman2001}.}
\end{center}

Figure~\ref{deformation} shows the equilibrium deformation
parameters $\beta_2$ and $\gamma$ obtained from the calculated TRS
minima for frequencies ranging from $\hbar$$\omega$ = 0.0 to 0.5 MeV
(The corresponding spin maximum of even-even $^{132-138}$Nd can be
extended up to about 20, 22, 16 and 14$\hbar$, respectively). Their
ground-state values are compared with other calculations and
experiments~\cite{muller1995,muller2008,Raman2001}, showing that our
results are close to the experimental values though there is still a
systematic underestimation for $\beta_2$. The negative $\beta_2$ of
$^{138}$Nd given by M\"{o}ller {\it et al}~\cite{muller1995}
indicates this nucleus is oblate, which is different with our
results and experiments. As expected, the ground-state $\beta_2$
deformations of these nuclei increase as the neutron number N moves
away from the closed shell N = 82. The calculated $\gamma$ values
are generally in agreement with the calculations by M\"{o}ller {\it
et al}~\cite{muller2008} except for that in $^{134}$Nd. Note that
the nucleus with $\pm\gamma$ deformations (e.g. in $^{138}$Nd) has
the same shape, as mentioned above. The $\beta_2$ deformations in
these nuclei almost keep constant as a function of rotational
frequency. The change of $\gamma$ value may provide the evolution
information of the triaxial shape and rotational axis. As shown in
Fig.1, it can be easily found that $^{132}$Nd has a prolate shape
with $\gamma = 0^\circ$. $^{134,136}$Nd exhibit the evolutions of
the rotational axes from the short-axis to medium-axis to
short-axis, moreover, $^{134}$Nd has a prolate collective rotation
beyond $\hbar\omega = 0.4MeV$. For $^{138}$Nd, the nucleus with
$\gamma \thickapprox -29^\circ$ rotates around the medium axis at
first and then a shape transition from the prolate-triaxial to
oblate-triaxial takes place at $\hbar\omega \thickapprox 0.2 MeV$,
the nucleus with $\gamma \thickapprox -84^\circ$ begins to rotate
around the long axis.

\begin{center}
\includegraphics[width=8cm]{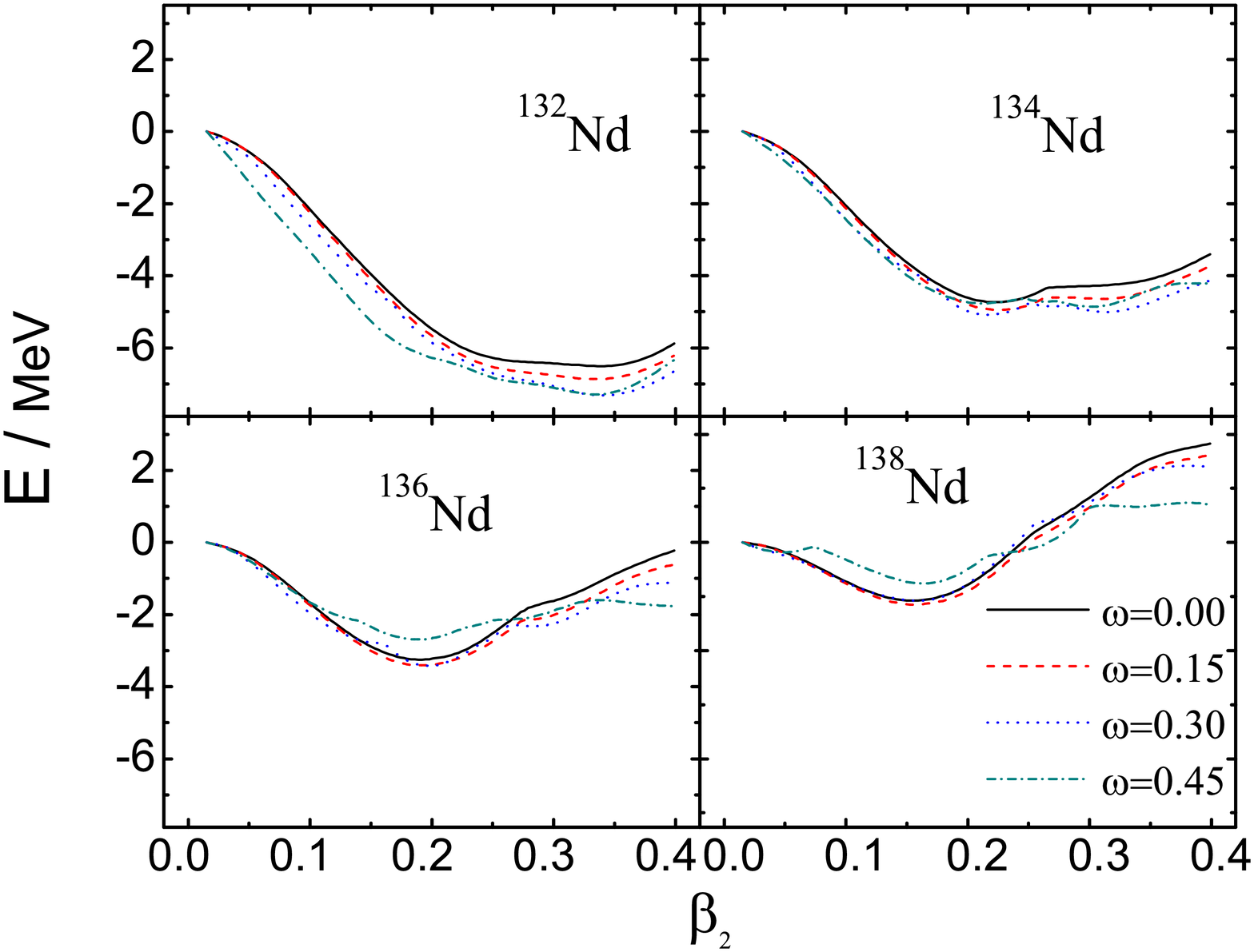}
\figcaption{\label{EBeta} Energy curves against $\beta_2$
deformation for even-even $^{132-138}$Nd nuclei at several selected
rotational frequencies $\omega$ = 0.00 (solid lines), 0.15 (dash
lines), 0.30 (dot lines) and 0.45 (dash-dot lines) $MeV/\hbar$. At
each $\beta_2$ point, the energy has been minimized with respect to
$\gamma$ and $\beta_4$.}
\end{center}

\begin{center}
\includegraphics[width=8cm]{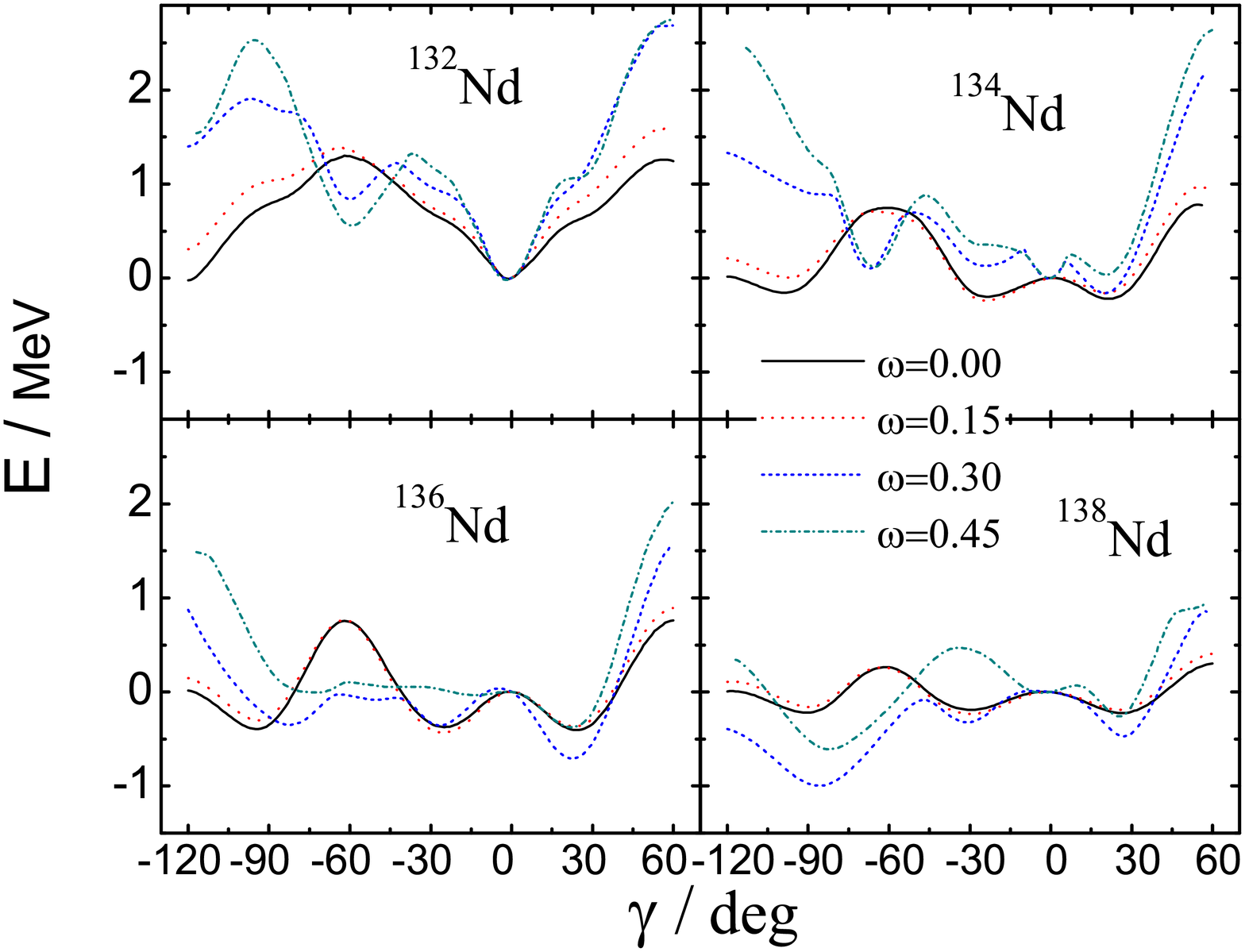}
\figcaption{\label{EGamma}  Similar to figure 2 but against $\gamma$
deformation.}
\end{center}

For transitional nuclei, the level scheme is more complicated than
that of spherical or well-deformed nuclear shape. Theoretical
studies are usually model-dependent because their equilibrium shapes
are generally soft and strongly affected by the mean-field and
pairing potential parameters. However, the rigidity or softness of
the nucleus, which can not be seen in Fig.~\ref{deformation}, is
almost independent of model parameters. To visually display the
nuclear softness in both $\beta_2$ and $\gamma$ directions, we show
the corresponding energy curves in Figs.~\ref{EBeta} and
 \ref{EGamma}. At each nucleus, four typical rotational frequencies
are selected to investigate the softness evolution with rotation.
One can see the rotational effects on the quadrupole deformation
$\beta_2$ are small, as shown in Fig.~\ref{EBeta}, implying that the
shapes are basically rigid against $\beta_2$ variation. On the
contrary, Figure~\ref{EGamma} shows that the energy curves as a
function of $\gamma$ deformation are strongly affected by the
cranking. At the ground states, the triaxial minima are rather
shallow. Theoretical and systematic studies indicate that the $N =
76$ nuclei are more $\gamma$-rigid than their neighbors with other
neutron number~\cite{Kortelahti1990,Kern1987}. It seems that the
depth of the triaxial minimum in $^{136}$Nd$_{76}$ is indeed
largest.  The evolution of the softness and depth of the minimum is
clearly presented under rotation, including that of the non-yrast
minimum. For instance, the prolate-obalte shape coexistence observed
in experiments~\cite{Petrache1997} can be found in $^{134}$Nd, as
shown in Fig.~\ref{EGamma}, and the similar situation exists in
$^{132}$Nd. In $^{136,138}$Nd, the coexistence of prolate-triaxial
and oblate-triaxial shapes is possible and awaits experimental
confirmation.

\begin{center}
\includegraphics[width=8cm]{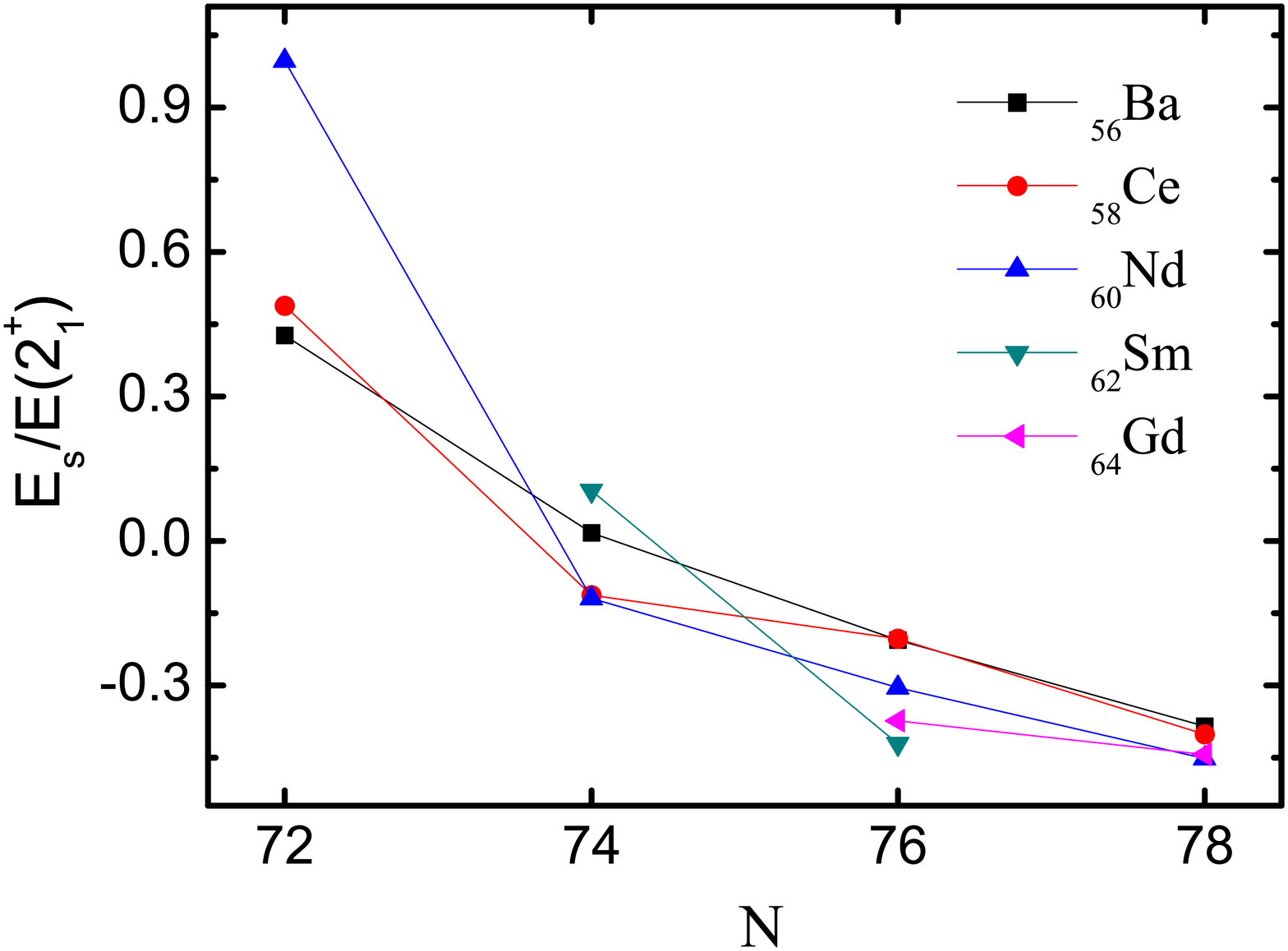}
\figcaption{\label{ES} Empirical values of the quantity
E$_s$/E(2$_1^+$) for even-even Nd and its adjacent Ba, Ce, Sm and Gd
isotopes versus neutron number. The available experimental data are
taken from
Ref.~\cite{Saito2008,Paul1987,Kortelahti1990,Angelis1994,Petrache2012,Li2013,NNDC}.}
\end{center}

The quantity $E_S/E(2^+_1)$, $E_S = E(2^+_2)- E(4^+_1 )$, can be as
a global signature of the structural evolution involving axial
asymmetry~\cite{Watanabe2011}. In the extreme $\gamma$-unstable
limit~\cite{Wilets1956}, the value of $E_S/E(2^+_1)$ is zero due to
the completely degenerate $2^+_2$ and $4^+_1$ states. In the case of
rigid-triaxial rotor with $25^\circ$
$\leqslant\gamma\leqslant30^\circ$~\cite{Davydov1958}, the $2^+_2$
state goes under the $4^+_1$ level and reaches the bottom at the
extreme of triaxiality with $\gamma=30^\circ$
($E_S/E(2^+_1)$=-0.67). Therefore, nuclei with negative values of
E$_s$/E(2$_1^+$) between these two extremes 0 and $-0.67$ are most
likely characterized by $\gamma$-soft potentials with shallow minima
at the average $\gamma$ value close to 30$^{\circ}$. Meanwhile, the
positive value of E$_s$/E(2$_1^+$) indicates that the nucleus
possesses an axially-symmetric shape, because the $2^+_2$ state lies
at high excitation energy relative to the $2^+_1$ and $4^+_1$
states.

The empirical E$_s$/E(2$_1^+$) values for even-even $^{132-138}$Nd
are presented in Fig.~\ref{ES}, showing also the results of their
adjacent nuclei $^{128-134}$Ba, $^{130-136}$Ce, $^{136,138}$Sm and
$^{140,142}$Gd. One can see a general decrease in $E_S/E(2^+_1)$
with increasing $N$, which illuminates the increasing of $\gamma$
softness in these isotopes. The observed rapid decrease from 0.99
($^{132}$Nd) to -0.45 ($^{138}$Nd) in $E_S/E(2^+_1)$ in the Nd
isotopic chain reflects the structural change from a nearly axial
rotor with the small-amplitude $\gamma$ vibrations to a
large-amplitude $\gamma$-soft dynamics. The $\gamma$-soft properties
of even-even $^{134-138}$Nd, as shown in Fig.~\ref{EGamma}, are also
supported by the negative $E_S/E(2^+_1)$ values. The $E_S/E(2^+_1)$
value in $^{138}$Nd is somewhat smaller than the empirical value of
$E_S/E(2^+_1)\thickapprox 0.5$, which is characteristic of the
critical-point nuclei in terms of maximum $\gamma$ softness between
prolate and oblate shapes. It should be noted that a critical point
of a prolate¨Coblate phase transition in $\gamma$-soft nuclei is
discussed in the context of the O(6) limit of the interacting boson
model, along with an interpretation in terms of Landau theory. This
is in good agreement with our calculated results of
$\gamma\thicksim29^\circ$ in ground state, as shown in
Fig.~\ref{deformation}. Our calculations show that the quantum phase
transition from triaxial-prolate to triaxial-oblate shapes at
$^{138}$Nd occurs at low-lying rather than ground states. The
calculations of M\"{o}ller {\it et al}~\cite{muller1995} indicate,
however, that  such phase transition has already taken place at
$^{138}$Nd ($\beta_2<0$). Figure~\ref{ES} also shows that $^{138}$Nd
with the lowest $E_S/E(2^+_1)$ value is the $\gamma$-softest nucleus
in this mass region. Indeed, $^{138}$Nd exhibits rather
$\gamma$-vibrational behavior experimentally, as demonstrated by the
observation of the properties expected for rotational bands built on
one-$\gamma$ and two-$\gamma$ -phonon states~\cite{Li2013}. This is
also supported by the study of Gizon {\it et al}~\cite{Gizon1978}
where it is deduced that a shape change across
$\gamma\approx30^\circ$ from prolate to oblate occurs between the
N=77 and N=79 Nd isotopes.

The energy staggering of the odd- and even-spin levels of a $\gamma$
band usually can be viewed as an important structural indicator to
distinguish between $\gamma$-rigid and $\gamma$-soft
asymmetry~\cite{Zamfir1991}. In such two different cases, though the
energies of the ground-state band are similar, the $\gamma$ band
nevertheless exhibits a different energy staggering. That is, the
$\gamma$-band levels of a rigid triaxial potential form couplets
arranged as ($2_{\gamma}^+$, $3_{\gamma}^+$), ($4_{\gamma}^+$,
$5_{\gamma}^+$), ($6_{\gamma}^+$, $7_{\gamma}^+$), $...$ while for a
completely $\gamma$-flat potential it has couplets ($2_{\gamma}^+$),
($3_{\gamma}^+$, $4_{\gamma}^+$), ($5_{\gamma}^+$, $6_{\gamma}^+$),
$...$ . Therefore, odd-even staggering in a $\gamma$ band can be
studied using the quantity~\cite{McCutchan2007}
\begin{equation}
  S(I) =
  \frac{E(I_{\gamma}^+)+E[(I-2)_{\gamma}^+]-2E[(I-1)_{\gamma}^+]}{E(2_1^+)},
\end{equation}
in which the energy differences are normalized to  $E(2_1^+)$.
Obviously, the sign of $S(I)$ is a strong indicator of the nature of
the $\gamma$ degree of freedom. For a rigid triaxial potential,
$S(I)$ will exhibit an oscillating behavior that takes on positive
and negative values for even and odd spins, respectively. In both
the vibrator and  limits, an opposite phase appears in $S(I)$,
namely positive for odd spin and negative for even spin. Moreover,
the overall magnitude of $S(I)$ is larger in the $\gamma$-soft limit
and increases gradually with spin compared with the vibrator
predictions that are smaller in magnitude and constant. For an
axially symmetric deformed rotor, the $S(I)$ values are positive,
small, and constant as a function of spin~\cite{McCutchan2007}.

\begin{center}
\includegraphics[width=8cm]{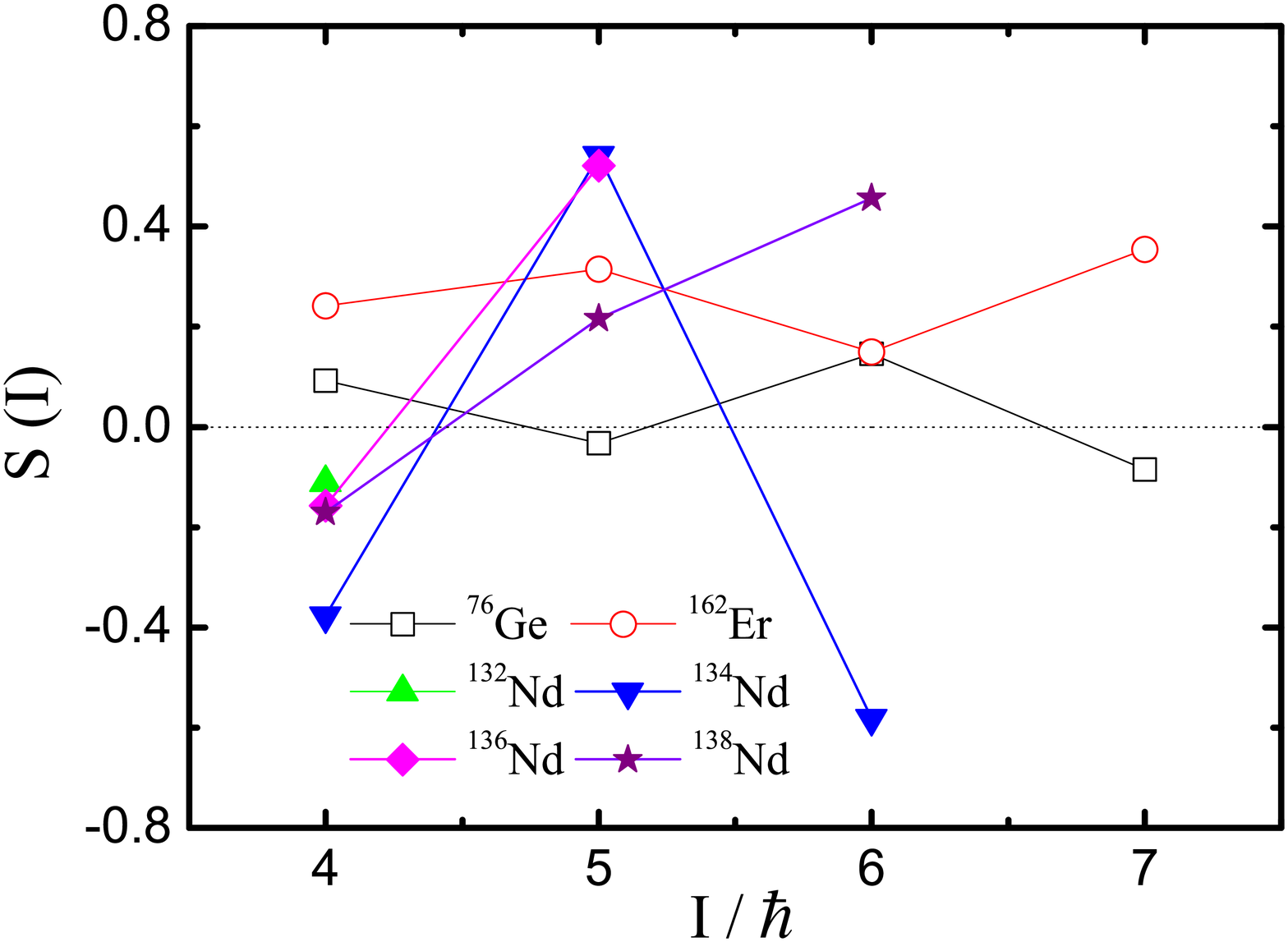}
\figcaption{\label{SI}   Odd-even energy staggering $S(I)$ for
even-even $^{132-138}$Nd nuclei, together with the $\gamma$-rigid
nucleus $^{76}$Ge and axially-symmetric nucleus $^{162}$Er for
comparison. These available data are from
Refs.~\cite{Paul1987,Kortelahti1990,Angelis1994,Petrache2012,Li2013,McCutchan2007,Toh2013}}
\end{center}

Experimental staggering $S(I)$ for even-even $^{132-138}$Nd nuclei
are shown in Fig.~\ref{SI} in comparison with those for a rigid
triaxial nucleus $^{76}$Ge~\cite{Toh2013} and an axially symmetric
nucleus $^{162}$Er~\cite{McCutchan2007}. It is unambiguous that the
sign of $S(4)$ is negative for the Nd isotopes. The magnitude of
$S(4)$ is smallest for $^{132}$Nd and largest for $^{134}$Nd. The
oscillatory pattern of $S(I)$ observed in $^{134}$Nd and at least
$S(4)$ and $S(5)$ in $^{136,138}$Nd, opposite to that in $^{76}$Ge,
agrees with the $\gamma$-soft potential predictions. The overall
magnitude of the staggering displays an increasing trend with spin
$I$ except for $^{132}$Nd in which only one $S(4)$ point is plotted
due to the scarce data. Especially for $^{136}$Nd, one can see a
rapidly increasing staggering which may indicates the $\gamma$
softness increases with increasing spin. It should be noted that if
the experimental data suggested in Ref.~\cite{Li2013} are used the
$S(6)$ value will be positive, which is not expected for a
$\gamma$-soft case. If this is actually the case, it will be
necessary to reveal the mechanism behind this anomalous behavior.
These properties are basically consistent with the
$\omega$-dependent energy curves, as shown in Fig.~\ref{EGamma}. To
investigate the evolution of $\gamma$ softness with rotation, it
will be of interest to further identify the high-spin levels of the
$\gamma$ band in future experiments.

\section{Summary}

In summary, doubly even $^{132-138}$Nd nuclei have been investigated
in terms of the TRS calculations in the ($\beta_2$, $\gamma$,
$\beta_4$) deformation space, focusing on the evolution of the
$\gamma$-softness and triaxiality with rotation. Compared to other
calculations and experiments, the equilibrium deformation parameter
$\beta_2$ and $\gamma$ are evaluated and it indicates our results
are closer to the experimental values. In these soft nuclei, the
existing differences of the equilibrium $\beta_2$ and $\gamma$
deformations between present work and other calculations may be
attributed to model parameters in a large extent. The nuclear
softness in the $\beta_2$ and $\gamma$ directions are displayed
using the corresponding energy curves at several selected rotational
frequencies. As the important structural indictors of axial
asymmetry, the quantities $S(I)$ and E$_s$/E(2$_1^+$) are analyzed
and discussed. The general trends are agreement with our
calculations. Meanwhile, it is pointed out that more detailed data,
especially in the $\gamma$ band, is needed to investigate and
confirm the $\gamma$-softness evolution in $^{132-138}$Nd. Also, it
should be noted that the present method does not include the effects
of rotation-vibration coupling, nor does include a
tilted-axis-cranking calculation, but can provide a qualitative
description of $\gamma$ correlations that are at least consistent
with some observed properties in some extent. A more reasonable
calculation, which is our future work, should take these effects
into account, especially for the transitional soft nuclei.

\end{multicols}
\vspace{-1mm} \centerline{\rule{80mm}{0.3pt}} \vspace{2mm}

\begin{multicols}{2}

\end{multicols}

\clearpage
%

\begin{thebibliography}{90}
\vspace{3mm}
\bibitem{Tajima2001}Tajima N, Suzuki N. Phys. Rev. C, 2001, {\bf 64}: 037301
\bibitem{Bengtsson1984}Bengtsson R, Frisk H, May F R et al. Nucl. Phys. A, 1984, {\bf 415}: 189-214
\bibitem{Meng1997}Frauendorf S, Meng J. Nucl. Phys. A, 1997, {\bf 617}: 131-147
\bibitem{Davydov1958}Davydov A S, Filippov G F. Nucl. Phys, 1958, {\bf 8}: 237-249
\bibitem{Wilets1956}Wilets L, Jean M. Phys. Rev, 1956, {\bf 102}: 788
\bibitem{Zamfir1991}Zamfir N V, Casten R F. Phys. Lett. B, 1991, {\bf 260}: 265
\bibitem{Nomura2012}Nomura K, Shimizu N, Vretenar D et al. Phys. Rev. Lett, 2012, {\bf 108}: 132501
\bibitem{Casten1985}Casten R F, Brentano P V, Phys. Lett, 1985, {\bf 152}: 22
\bibitem{Chen1983}CHEN Y S, Frauendorf S, Leander G A. Phys. Rev. C, 1983, {\bf 28}: 2437
\bibitem{Saito2008}Saito T R, Saito N, Starosta K et al. Phys. Lett. B, 2008,  {\bf 669}: 19¨C23
\bibitem{Wang20121}WANG H L, LIU H L, XU F R et al. Prog. Theo. Phys., 2012, {\bf 128}: 363-371
\bibitem{Wang20122}WANG H L, LIU H L, XU F R. Phys. Scr., 2012, {\bf 86}: 035201
\bibitem{Petrache1996}Petrache C M, Bazzacco D, Lunardi S et al. Phys. Lett. B, 1996, {\bf 387}: 31-36
\bibitem{Mukhopadhyay2008}Mukhopadhyay S, Almehed D, Garg U et al. Phys. Rev. C, 2008, {\bf 78}: 034311
\bibitem{Petrache1999}Petrache C M, Bianco G L, Ward D et al. Phys. Rev. C, 1999, {\bf 61}: 011305(R)
\bibitem{Wadsworth1988}Wadsworth R, O'Donnell J M, Watson D L et al. Nucl. Phys., 1988, {\bf 14}: 239-251
\bibitem{Angelis1994}Angelis G D, Cardona M A, Poli M D et al. Phys. Rev. C, 1994, {\bf 49}: 2990
\bibitem{Paul1987}Paul E S, Beausang C W , Fossan D B et al. Phys. Rev. C, 1987, {\bf 36}: 1853
\bibitem{Kortelahti1990}Kortelahti M O, Kern B D, Braga R A et al. Phys. Rev. C, {\bf 42}: 1267
\bibitem{Petrache2012}Petrache C M, Frauendorf S, Matsuzaki M et al. Phys .Rev. C, 2012, {\bf 86}: 044321
\bibitem{Li2013}LI H J, XIAO Z G, ZHU S J et al. Phys. Rev. C, 2013, {\bf 87}: 057303

\bibitem{Satula1994}Satu{\l}a W, Wyss R, Magierski P. Nucl. Phys. A, 1994, {\bf 578}: 45-61
\bibitem{Xu2000}Xu F R, Satu{\l}a W and Wyss R. Nucl. Phys. A, 2000, 669: 119.
\bibitem{Myers1966}Myers W D, Swiatecki W J. Nucl. Phys., 1966, {\bf 81}: 1-60
\bibitem{Cwiok1987}$\rm\acute{C}$wiok S, Dudek J, Nazarewicz W et al. Comp. Phys. Comm., 1987, {\bf 46}: 379-399
\bibitem{Greiner1996}Greiner W, Maruhn J A. \emph{Nuclear Models}. Springer-Verlag, 1996, 108
\bibitem{Naza1989}Nazarewicz W, Wyss R, Johnson A. Nucl. Phys., 1989, {\bf A503}: 285-330
\bibitem{Strutinsky1967}Strutinsky V M. Nucl. Phys., 1967, {\bf A95}: 420-442
\bibitem{Pradhan1973}Pradhan H C, Nogami Y, Law J. Nucl. Phys. A, 1973, {\bf A201}: 357-368
\bibitem{Moller1992}M$\rm\ddot{o}$ller P, Nix J R. Nucl. Phys. A, 1992, {\bf A536}: 20-60
\bibitem{Sakamoto1990}Sakamoto H, Kishimoto T. Phys. Lett. B, 1990, {\bf 245}: 321

\bibitem{muller1995}M$\rm\ddot{o}$ller P, Nix J R, Myers W D et al. At. DATA Nucl. Data Tables, 1995, {\bf 59}: 275
\bibitem{muller2008}M$\rm\ddot{o}$ller P, Bengtsson R, Carlsson et al. At. DATA Nucl. Data Tables, 2008, {\bf 94}: 770
\bibitem{Raman2001}Raman S, Nestor C W, JR. et al. At. DATA Nucl. Data Tables, 2001, {\bf 78}: 43-44
\bibitem{Kern1987}Kern B D, Mlekodaj R L, Leander G A et al. Phys. Rev. C, 1987, {\bf 36}: 1514
\bibitem{Petrache1997}Petrache C M, Sun Y, Bazzacco D et al. Nucl. Phys. A, 1997, {\bf 617}: 249
\bibitem{Watanabe2011}Watanabe H, Yamaguchi K, Odahara et al. Phys. Lett. B, 2011, {\bf 704}: 270-275
\bibitem{NNDC} http://www.nndc.bnl.gov/
\bibitem{Gizon1978}Gizon J, Gizon A, Diamond R M et al. Nucl. Phys., 1978, {\bf 4}: L171
\bibitem{McCutchan2007}McCutchan E A, Bonatsos D, Zamfir N V et al. Phys. Rev. C, 2007, {\bf 76}: 024306
\bibitem{Toh2013}Toh Y, Chiara C J, McCutchan E A et al. Phys. Rev. C, 2013, {\bf 87}: 041304(R)


\end{thebibliography}
\end{document}